\documentclass[12pt]{article}
\usepackage{graphicx}
\usepackage{cite}
\newcommand{\dissum}[2]{\displaystyle \sum_{#1}^{#2}}
\newcommand{\fnd}[2]{\frac{\textstyle #1}{\textstyle #2}}

\newcommand{\abs}[1]{\left| #1\right|}
\newcommand{\babar}
{{\it B}$\!${\footnotesize\it A}$\!${\it B}$\!${\footnotesize\it A$\!$R}}
\begin{document} \baselineskip .7cm
\title{Threshold behaviour of scattering poles}
\author{
Eef van Beveren\\
{\normalsize\it Centro de F\'{\i}sica Te\'{o}rica}\\
{\normalsize\it Departamento de F\'{\i}sica, Universidade de Coimbra}\\
{\normalsize\it P-3004-516 Coimbra, Portugal}\\
{\small eef@teor.fis.uc.pt}\\ [.3cm]
\and
George Rupp\\
{\normalsize\it Centro de F\'{\i}sica das Interac\c{c}\~{o}es Fundamentais}\\
{\normalsize\it Instituto Superior T\'{e}cnico, Edif\'{\i}cio Ci\^{e}ncia}\\
{\normalsize\it P-1049-001 Lisboa Codex, Portugal}\\
{\small george@ajax.ist.utl.pt}\\ [.3cm]
{\small PACS number(s):  11.80.Et, 12.40.Yx, 13.75.Lb, 14.40.-n} \\[1.0cm]
{\normalsize Talk given at the}\\
{\normalsize High-Energy Physics Workshop}\\
{\normalsize\it Scalar Mesons: an Interesting Puzzle for QCD}\\
{\normalsize May 16 - 18, 2003
}\\
{\normalsize SUNY Institute of Technology, Utica (NY)} \\[0.3cm]
{\small hep-ph/0306185}
}

\maketitle

\begin{abstract}
The results of a model for meson-meson scattering are studied.
The model is shown to be capable of on the one hand reproducing the scattering
data, while on the other hand a quark-antiquark confinement spectrum can be
determined.

It is concluded that adopting the model's formulation of the transition matrix
elements for data analysis , it may serve as a link between experiment and
quenched lattice calculations.
\end{abstract}

\section{Introduction}

\begin{figure}[htbp]
\begin{center}
\begin{tabular}{cc}
\resizebox{6.0cm}{!}{\includegraphics
{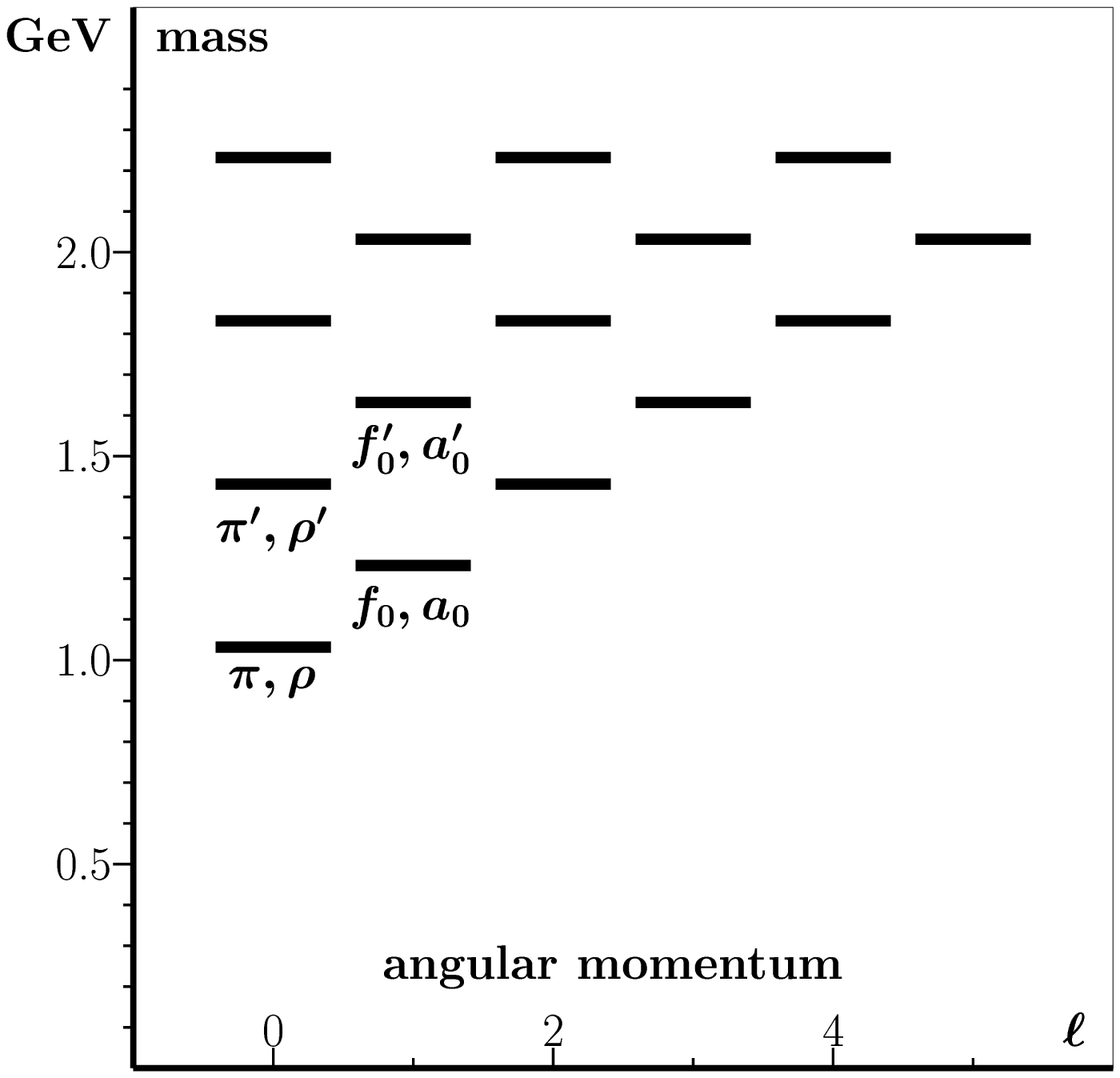}} &
\resizebox{6.0cm}{!}{\includegraphics
{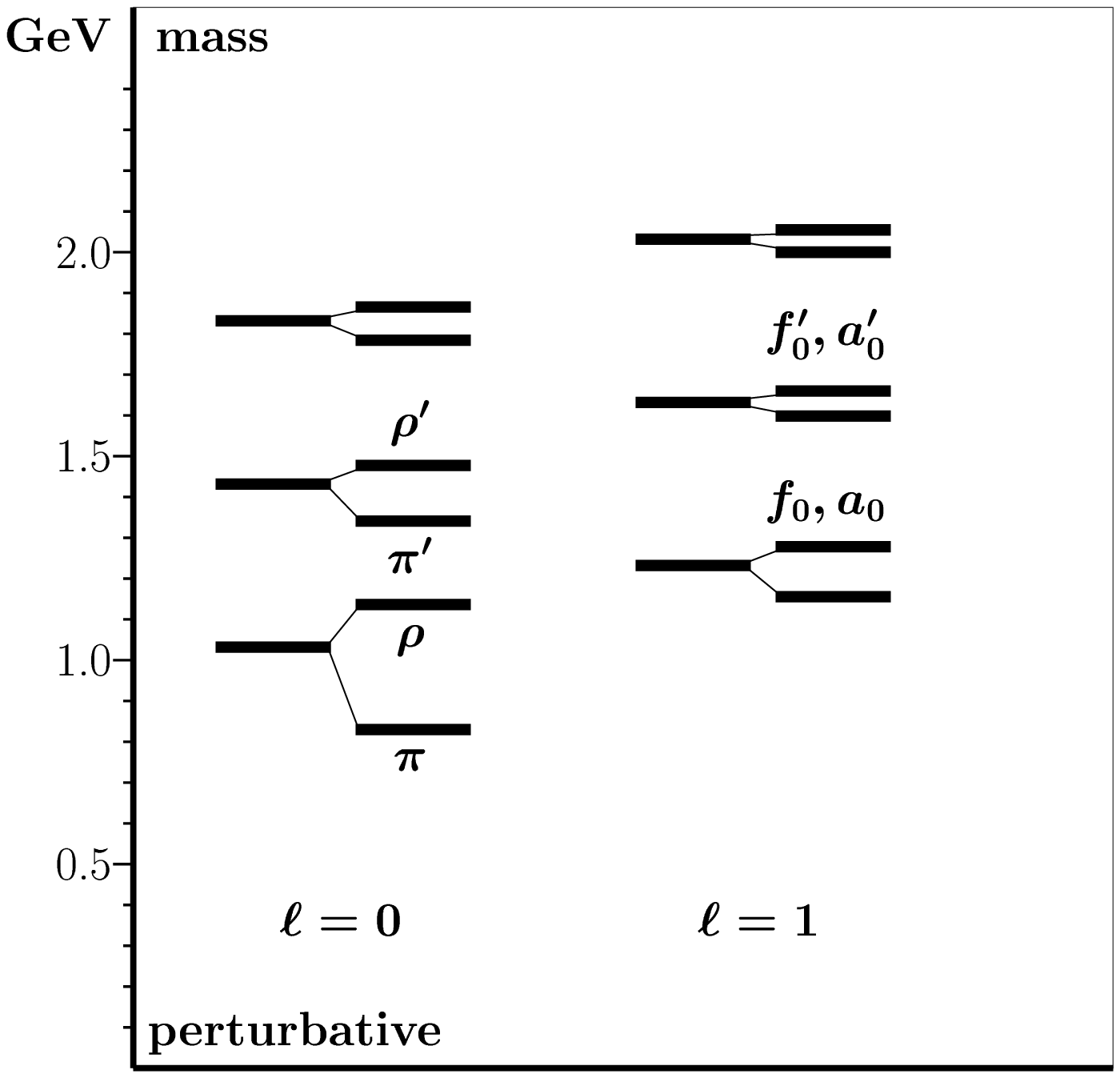}}\\
(a) & (b)
\end{tabular}
\end{center}
\normalsize
\caption{Mass spectrum for a permanently closed system
without (a), and including short distance effects perturbatively (b).
The parameters are taken from Ref.~\cite{PRD27p1527}, where harmonic-oscillator
confinement is employed.}
\label{HO}
\end{figure}

In the sixties, early seventies, mesons and baryons were studied in terms
of models for confinement of the newly invented quarks
\cite{CERNREPTH401/412,PR125p1067}.
The basic idea was that mesons and baryons could be described by
permanently confined quarks and/or antiquarks.
In Fig.~(\ref{HO}a) we give an artist's impression of the resulting mass
spectrum for mesons.

\begin{figure}[htbp]
\begin{center}
\begin{tabular}{cc}
\resizebox{6.0cm}{!}{\includegraphics
{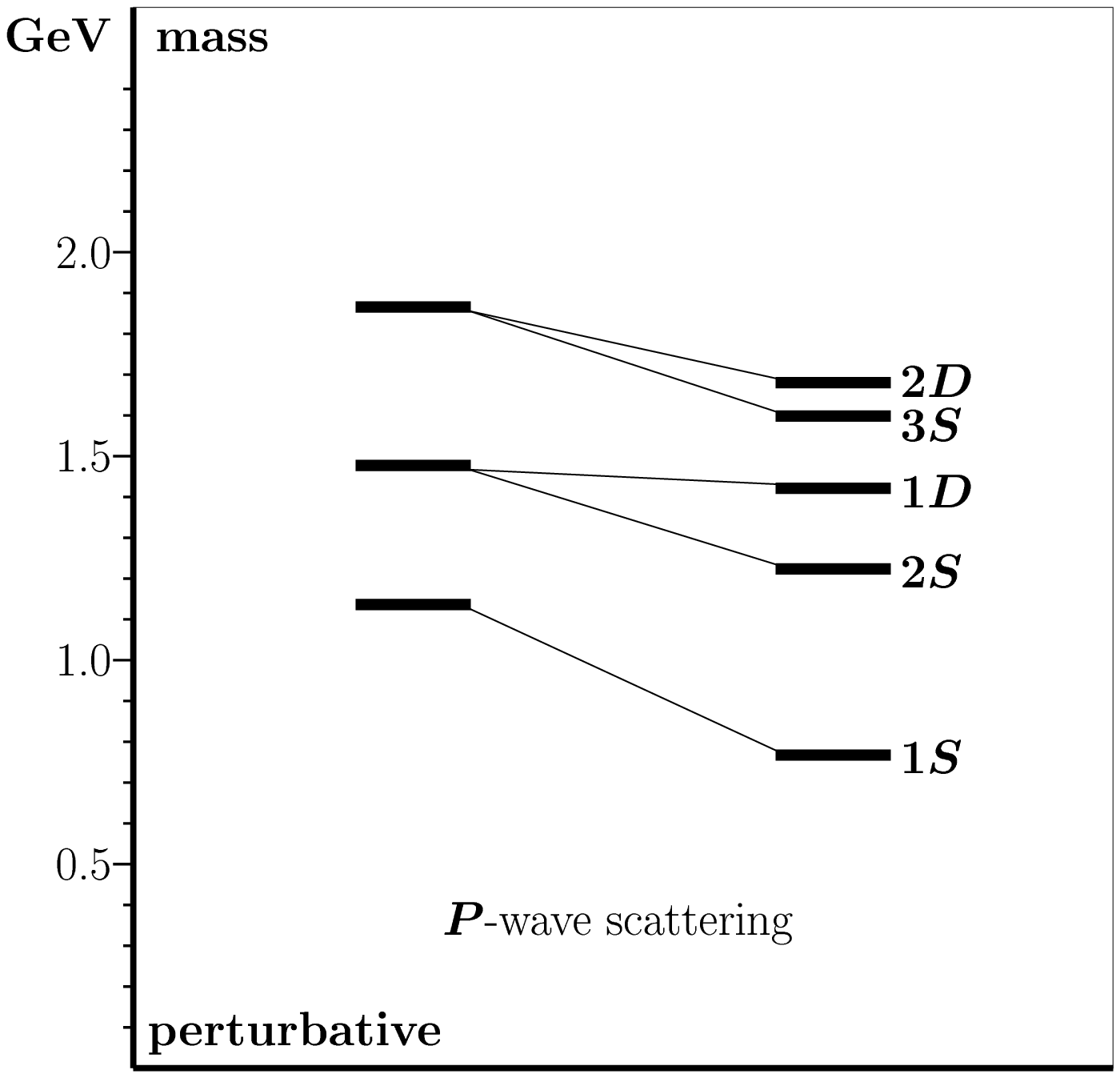}} &
\resizebox{6.0cm}{!}{\includegraphics
{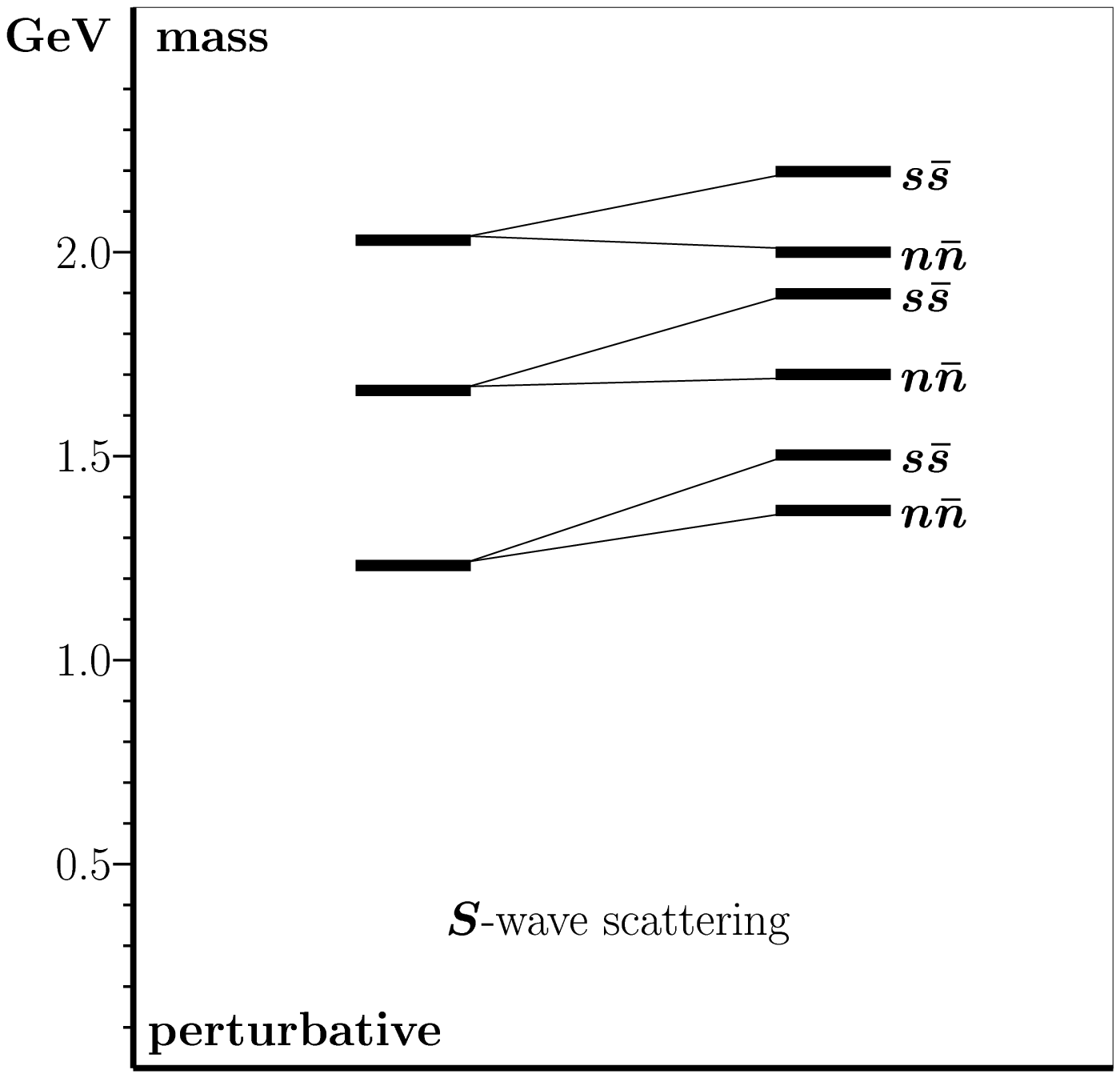}}\\
(a) & (b)
\end{tabular}
\end{center}
\normalsize
\caption{Mass splittings for a permanently closed system when hadronic
decay is perturbatively taken into account,
for $P$ wave (a), and for $S$ wave (b).}
\label{decay}
\end{figure}

Then, in the seventies, early eighties, short-range effects had to overcome
the small disagreements between Nature and confinement models, as
schematically depicted in Fig.~(\ref{HO}b).
Later, hadronic decay was implemented, with effects on the meson masses
as shown in Fig.~\ref{decay}.

However, it was all the time overlooked how those mesons are actually
produced in experiment.
In particular, it was taken for granted that mesonic states below
a particular meson-meson threshold do not feel the presence of that
two-meson channel.
Effects of coupling were assumed to abruptly vanish or show up,
depending on whether the state was just above or just below threshold.
For states far below threshold, the whole idea of implementing
two-meson channels in the confinement models was considered
inopportune.
Furthermore, little attention was paid to the fact that most mesons
appear as resonances, and that mesonic resonance widths are usually not small.
For example, a $\rho$ meson has a lifetime compatible with its
internal frequency, implying that it should certainly not be considered
a more or less stable particle.
Even nowadays this is not fully accepted or even understood.

The procedure that mimics the experimental situation 
is called {\it unitarization} \/at present.
This name suggests that all is accounted for, which is of course an
illusion.
In practice, one is more than satisfied if the ``most important'' effects
are included in the resulting meson model.

Also in the seventies one became aware of phenomena
in $S$-wave meson-meson scattering which could not be easily handled by any
of the models.
Hence, still in the spirit of permanent confinement,
quark configurations were invented for mesons and baryons other
\cite{PRD15p267,PRD21p1370,PRD21p2653} than
the usual $q\bar{q}$ and $q^{3}$.

However, things really fall in place when full scattering properties
are determined for meson-meson, or meson-baryon, processes,
since then resonances come out automatically without any need to worry about
their composition.
\vspace{0.2cm}

We observe the following:
\vspace{0.2cm}

\begin{itemize}
\item The higher radial excitations in $c\bar{c}$ and $b\bar{b}$ vector states
are almost equally spaced in mass.
Why the ground states for those systems come out much lower
is well explained \cite{PRD21p772}.
\item The level splittings for the $S$ and $D$ $c\bar{c}$ and $b\bar{b}$
vector states follow naturally \cite{PRD21p772}.
\item The phase shifts for $K\pi$ and $\pi\pi$ in $P$ wave \cite{PRD27p1527},
and in $S$ wave \cite{ZPC30p615}, as well as the scattering lengths
\cite{EPJC22p493} are reproduced.
\item The $J^{P}=0^{+}$ $c\bar{s}$ experimental results are explained
\cite{HEPPH0305035}.
\item The $J^{P}=1^{+}$ $c\bar{n}$ ($n$ for nonstrange, either up or down)
and $c\bar{s}$ phenomena come out well \cite{HEPPH0306051}.
\item It fully explains the light scalar meson nonet \cite{ZPC30p615}.
\end{itemize}
\vspace{0.2cm}

No extra forces and/or configurations are necessary to explain
such an amount of very different data, with one set of parameters for all
mesons.

We may thus conclude that unitarization works well.
Nevertheless, our model is far from perfect for many reasons that are
easy to understand.
First, for the transition potential,
which should have been derived from the theory of strong interactions,
we make an as simple as possible choice, and then it is still further
simplified \cite{syracuse}.
Furthermore, also harmonic-oscillator confinement does not have much more
justification than the equal-level spacings for $c\bar{c}$ and $b\bar{b}$
(and possibly also $n\bar{n}$ and $s\bar{s}$) vector states.
Hence, our choices for both, the confinement and unitarization ingredients,
are taken as simple as possible.
The fact that it works even under such extreme simplifications clearly
indicates that, in a more elaborate approach, the unitarization scheme has a
good chance to survive all tests.

\section{Unitarization}

In order to understand our approach to unitarization,
one may imagine a huge scattering or transition matrix,
describing all possible scattering of meson pairs.
Elastic processes appear on its diagonal,
whereas off-diagonal matrix elements describe inelastic processes.
Conservation laws predict that the vast majority of off-diagonal matrix
elements vanish.
The remaining nonvanishing matrix elements can be regrouped into smaller
submatrices of meson-meson channels which under the given conservation
laws are allowed to communicate.
Hence, if we for example study the $J^{P}=0^{+}$ $c\bar{s}$ system,
then we consider the part of the $T$ matrix which describes the
$S$-wave scattering of all meson pairs that through OZI-allowed processes
couple to $c\bar{s}$.
In practice, we limit ourselves to a few channels which we ``believe''
are relevant to the energy domain under study.
In the following we shall concentrate on just one scattering channel.
But the results are equally valid when more channels are taken into account.

Let us now consider an arbitrary diagonal element of the above-discussed
transition matrix in a specific partial wave, which we denote by
$T_{\ell}$.
It describes the elastic scattering of two mesons, assumed not to couple to
anything else but the confinement system having the corresponding
quantum numbers.
We assume here that confinement yields an infinite spectrum.
The function $T_{\ell}$ must be analytic in the total invariant mass
$E=\sqrt{s}$ of the two-meson system.
Moreover, it depends on the meson masses $M_{1}$ and $M_{2}$,
the intensity $\lambda$ of the coupling between the
two-meson system and the confined quark-antiquark system,
and finally on the confinement spectrum
$E_{n}$ ($n=0$, 1, 2, $\ldots$) and other details ${\cal F}_{q\bar{q}}$
of the $q\bar{q}$ system.
In Ref.~\cite{syracuse}, the interested reader may find a rather general and
complete expression for $T_{\ell}$.
Summarizing, we conclude

\begin{equation}
T_{\ell}\; =\; T_{\ell}\left(\sqrt{s},\lambda ,M_{1},M_{2};
\left\{ E_{n}\, |\, n=0,\, 1,\, 2,\,\dots\,\right\},\dots\right)
\;\;\; .
\label{Wilhelm}
\end{equation}

\begin{figure}[htbp]
\centerline{\scalebox{0.7}{\includegraphics
{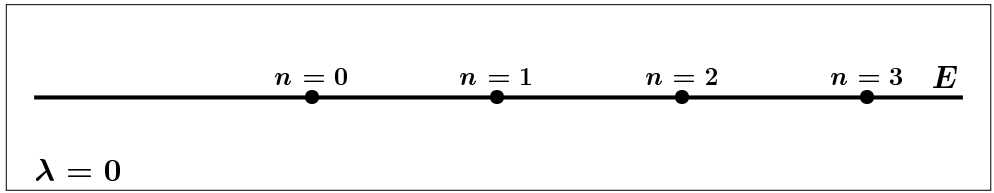}}}
\normalsize
\caption{The spectrum of confinement.}
\label{HOagain}
\end{figure}

In Fig.~\ref{HOagain} we have depicted a possible spectrum for confinement.
The spectrum corresponds to $\lambda=0$, i.e., having discrete energy
eigenvalues of the confinement operator under the assumption that the
$q\bar{q}$ system is decoupled from the meson-meson continuum.

\begin{figure}[htbp]
\centerline{\scalebox{0.95}{\includegraphics
{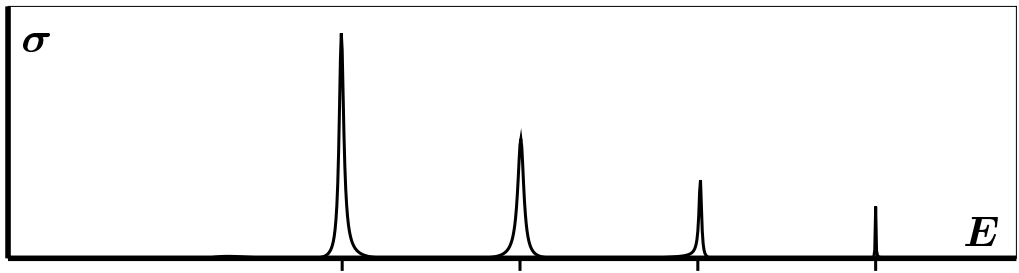}}}
\normalsize
\caption{Confinement spectrum as observed in  elastic meson-meson
scattering, for the case that the confinement states are only weakly
coupled to the meson-meson continuum.}
\label{HOpeaks}
\end{figure}

When the confinement states are weakly coupled to the meson-meson continuum
($\lambda$ small),
we observe narrow resonances in the two-meson scattering cross
section $\sigma$, about at the energy eigenvalues of confinement.
This is depicted in Fig.~\ref{HOpeaks}.

\begin{figure}[htbp]
\centerline{\scalebox{0.95}{\includegraphics
{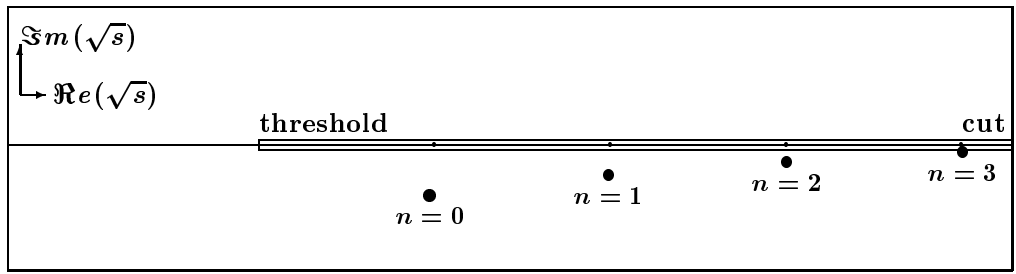}}}
\normalsize
\caption{The scattering-matrix poles associated with the resonances shown in
Fig.~\ref{HOpeaks}. Threshold is at $\sqrt{s}=M_{1}+M_{2}$.
The confinement energy eigenvalues are indicated by small dots
on the real-energy axis.}
\label{Polesagain}
\end{figure}

The cross section in Fig.~\ref{HOpeaks} is calculated by the use of the
transition matrix element $T_{\ell}$ of Eq.~~(\ref{Wilhelm}), an explicit
expression of which can be found in Ref.~\cite{syracuse}.
Now, since $T_{\ell}$ is an analytic function in the total invariant
two-meson mass $E=\sqrt{s}$, one may study its singularity structure for
complex $E$. This can be done numerically.
In Fig.~\ref{Polesagain}, we show the poles associated with the resonances
in the cross section depicted in Fig.~\ref{HOpeaks}.

The real parts of the poles are close to the energy eigenvalues
of the confinement spectrum, while the imaginary parts are relatively small as
the resonances are narrow.

In the following we study how the scattering-matrix poles move through
the complex-energy plane, when the coupling $\lambda$
between the meson-meson and confinement systems is increased,
in particular near threshold.

But first, let us discuss how a pole of the scattering matrix behaves
when the confinement spectrum ($\lambda=0$) has a state which is
below threshold.
In our model, these poles shift to even lower energies when $\lambda$
is increased \cite{PRD21p772}, but remain on the real-energy axis.
Such poles correspond to bound states of the coupled system.
In this case, both the two-meson and the $q\bar{q}$ components
of the wave function describe bound systems.
The corresponding states are thus mixtures of bound
two-meson and $q\bar{q}$ states.

\section{Behavior at threshold}

\begin{figure}[htbp]
\centerline{\scalebox{1.0}{\includegraphics
{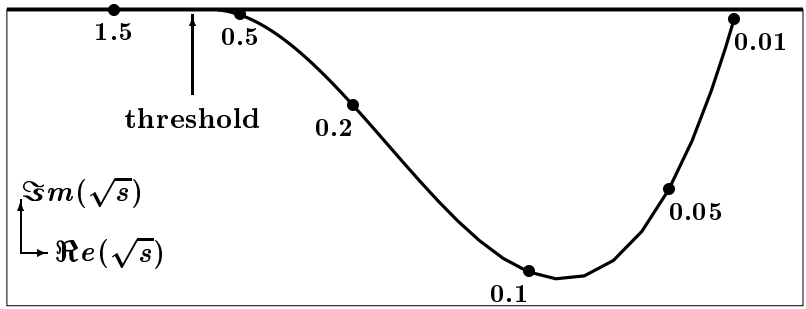}}}
\normalsize
\caption{Scattering-matrix pole position of the ground-state resonance,
as a function of the coupling constant $\lambda$,
for $P$- and higher-wave scattering.
Some values of $\lambda$ are indicated along the curve.}
\label{passthrough}
\end{figure}

When the ground state of the confinement spectrum is near a non-$S$-wave
threshold, then the associated $T$-matrix pole passes through threshold when
$\lambda$ is increased \cite{HEPPH0306051}.
This phenomenon is depicted in Fig.~\ref{passthrough}.

We observe that, while the real part of the pole is continuously decreasing
when $\lambda$ is increased, the imaginary part first increases
(in modulus), then decreases, vanishes right at threshold,
and remains zero when $\lambda$ is still further increased.
For small values of $\lambda$, the displacements of the scattering-matrix
poles in the complex-energy plane are almost perturbative,
i.e., linear in $\lambda^{2}$.
However, when $\lambda$ grows it becomes clear that the model
incorporates highly nonperturbative effects.

The pole displacement depicted in Fig~\ref{passthrough} represents
two-meson scattering in $P$ wave.
For higher waves the behavior is similar.
This can easily be understood by studying the effective-range expansion
near threshold.
For $P$ wave and higher waves, one finds that the imaginary part of the pole
is quadratic in $\lambda$ when the real part of the linear momentum is small,
which is near threshold.
Consequently, at threshold it vanishes.

However, $S$-wave poles behave very differently
\cite{AIPCP660p353,HEPPH0304105}. They approach the real-energy axis
perpendicularly well below threshold for increasing $\lambda$.
Then, for still larger values of $\lambda$, they first move in the positive
direction towards threshold, as virtual bound states.
At threshold they turn back, now moving towards smaller energies as genuine
bound states.
This can more easily be understood from their movement in the
complex-linear-momentum plane \cite{LNP211p331,AIPCP660p353}.
Experimentally, one can determine whether the pole represents
a virtual or a real bound state from the sign of the scattering length.

\section{Poles from the quark-pair-creation cavity}

For small coupling, it is very clear how poles in the scattering amplitude
are related to the energy eigenvalues of confinement,
since their displacements are perturbative, hence small.
However, for strong coupling pole displacements can be of the order of
magnitude of the level splittings of the confinement spectrum.
In thit case the situation is more complicated,
and were it not for a model to trace the poles,
the classification in terms of a confinement spectrum might turn impossible.

But there is more:
even when all poles are traced, starting from the small-$\lambda$ positions
near the confinement spectrum, and ending up at the physical positions fitting
the scattering data, there exist still more poles in the complex-energy plane.
This means that the scattering amplitude which agrees with experiment contains
more poles than just those stemming from the confinement spectrum.
But let us first discuss where such {\it extra} \/poles happen to originate
in the model.

In our model, the two-meson system communicates with the confined $q\bar{q}$
system through OZI-allowed quark-pair creation. The shape of the corresponding
potential has a maximum at relatively large distances (0.5--1.0 fm).
This means that in the interior one has a small potential well which in
principle could host bound states.
For small coupling, the well is almost flat and thus of no consequence,
giving rise to merely mathematical poles with very large negative imaginary
parts, hence unobservable.
However, for strong coupling these {\it cavity poles} \/will turn out to get
mixed up with the poles originating in the confinement spectrum.
It such a case, there are clearly observable effects in the scattering
cross section.
Under certain conditions, the poles even end up on the real-energy axis
as bound states. In the past we referred to these poles as {\it background}
\/poles, since for decreasing coupling they disappear into the background.
Here, we shall stick to the term {\it extra} \/poles.

The extra poles were first reported in Ref.~\cite{ZPC30p615}.
Later, it was thought that they were the result of pole
doubling \cite{ZPC68p647}, a term which we adopted for a while.
But now it has become clear from which mechanism they originate.

For meson-meson scattering in $S$ wave, it is more likely that certain
structures in the cross section stem from the {\it extra} \/poles
than for $P$ and higher waves.
The reason is that in $S$-wave scattering the centrifugal barrier is
absent.
Consequently, the transition-potential well is deeper in the latter case.
This is exactly what is observed in experiment.

\begin{figure}[htbp]
\begin{center}
\begin{tabular}{ccc}
\resizebox{5.3cm}{!}{\includegraphics
{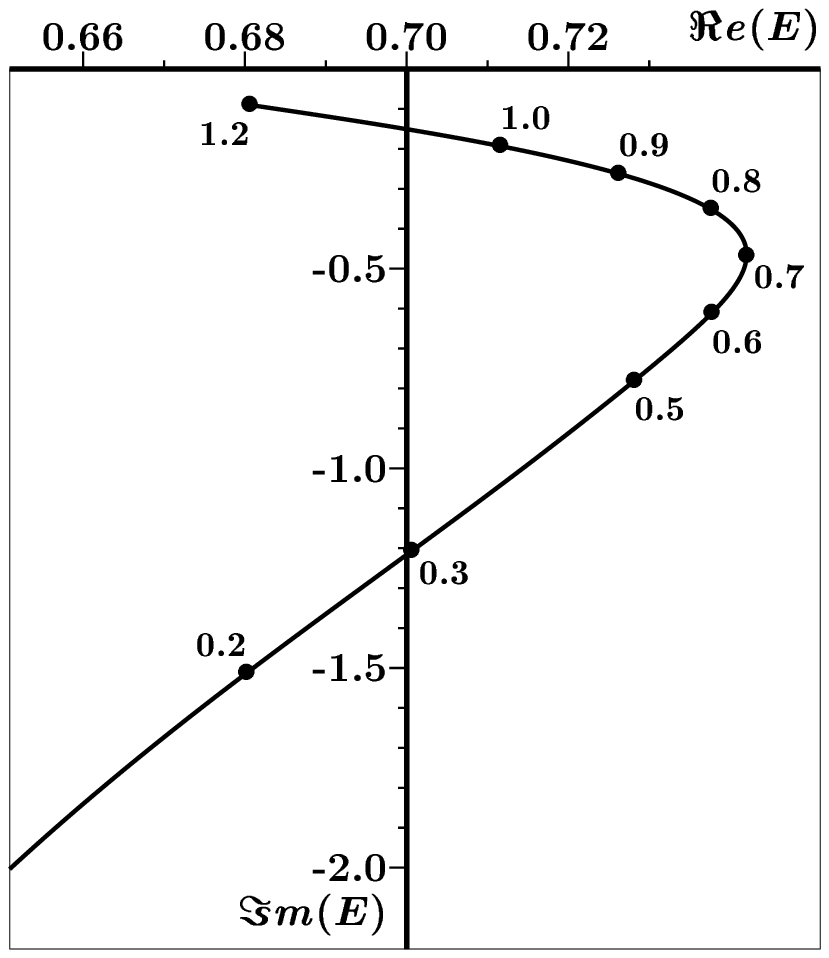}} &
\resizebox{5.3cm}{!}{\includegraphics
{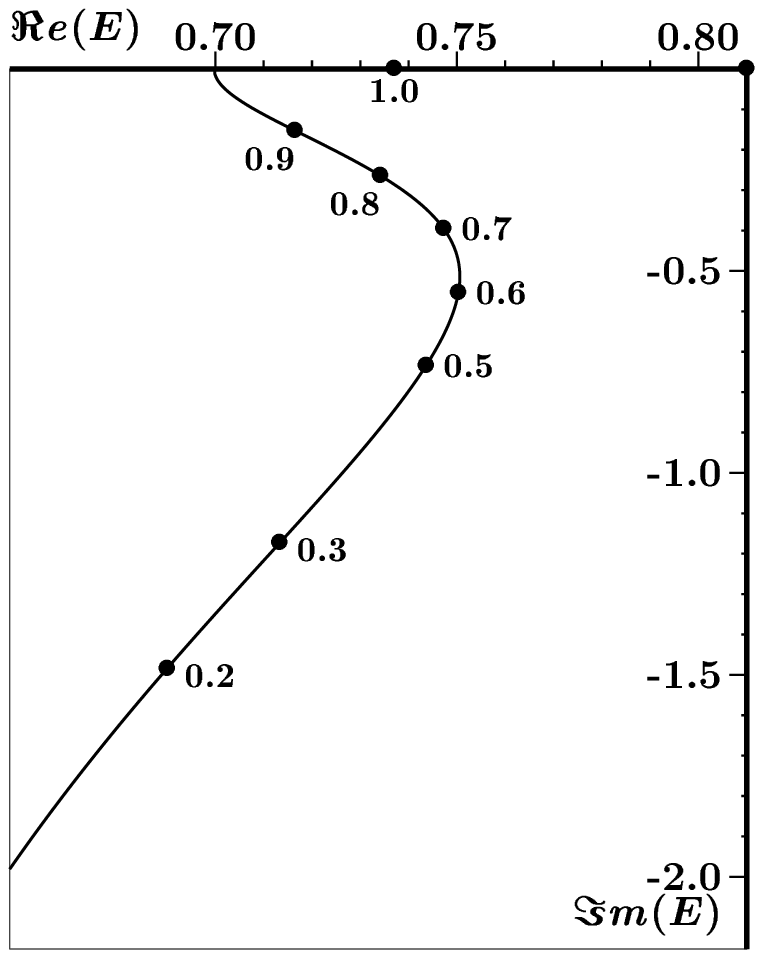}} &
\resizebox{5.3cm}{!}{\includegraphics
{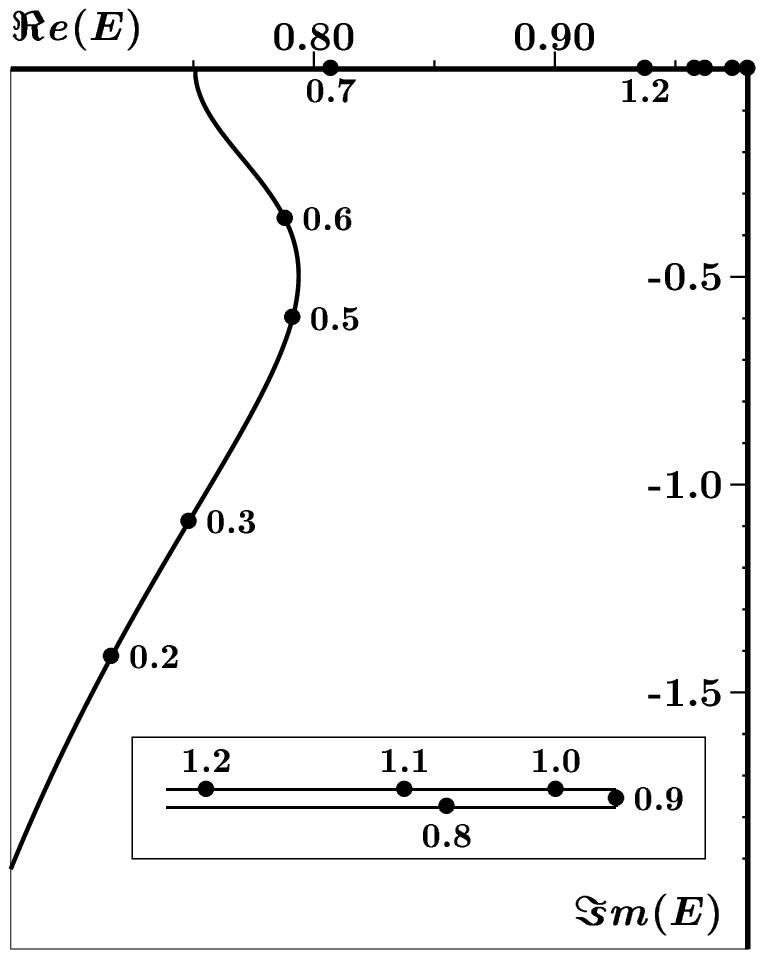}}\\
(a) & (b) & (c)
\end{tabular}
\end{center}
\normalsize
\caption{The extra poles w.r.t. threshold.}
\label{extrapole}
\end{figure}

There are several parameters influencing the displacements of
the extra poles, one of them being the relative position of threshold
with respect to the confinement spectrum.
When threshold is far below the confinement-spectrum ground state,
which is usually the case when pions are involved, then the extra pole
does not come close to the real-energy axis, causing a broad structure
in the scattering cross section, which often is not even a clear resonance.
When threshold is closer to the confinement-spectrum ground state,
then the extra pole comes closer to the real-energy axis,
but the real part of the pole position is below threshold,
causing structure at threshold.  When threshold is sufficiently close
to the confinement-spectrum ground state,
then the extra poles end up on the real-energy axis below threshold,
where they give rise to bound states, or possibly virtual bound states
close to threshold.
In Fig.~\ref{extrapole} we show the various possibilities.

In order to study the pole displacements as a function of the coupling,
we normalize $\lambda$ such that $\lambda =1$ corresponds to the physical
situation in Fig.~\ref{extrapole}.
The confinement spectrum is chosen to have a ground state at 1.3 GeV and
a level spacing of 0.4 GeV.
We study meson-meson scattering in $S$ wave.
One of the two mesons has a fixed mass of 0.5 GeV, the other is varied
in order to obtain different values for threshold.
A first observation from Fig.~\ref{extrapole} is that the extra pole resides
at minus infinity imaginary energy when the meson-meson system is uncoupled
from the confinement states.

From Fig.~\ref{extrapole}a, where the second meson mass is equal to 0.2 GeV,
so threshold is at 0.7 GeV, we observe that the extra pole comes out
at $(0.711-i0.20)$ GeV for $\lambda =1$.
In the cross section this gives rise to a very broad resonance-like
structure.

In Fig.~\ref{extrapole}b we choose 0.31 GeV for the second meson mass,
so threshold lies at 0.81 GeV.
In this case we find the physical pole on the real-energy axis,
some 73 MeV below threshold, but still before having reached threshold.
Such a pole represents a virtual bound state, which reflects itself in
some structure at threshold in the scattering cross section.
Only for $\lambda =1.2$ the pole arrives at threshold.
Notice also that the pole reaches the real-energy axis at 0.70 GeV,
which is 110 MeV below threshold.
Such a phenomenon only occurs for $S$-wave scattering.

In Fig.~\ref{extrapole}c the second meson mass equals 0.48 GeV,
which brings threshold to 0.98 GeV.
The situation near threshold is now more confusing, but we have enlarged
that part of the figure in the inset.
From the inset of Fig.~\ref{extrapole}c we learn that the physical pole
comes out some 10 MeV below threshold, but this time as a real bound state.
Note that in this case the pole arrives at the real axis 230 MeV below
threshold.

In the next section we discuss well-known experimental facts supporting
the existence of extra poles in the scattering amplitude for strong
interactions.

\section{The scalar mesons}

As extra poles show u  mainly in $S$-wave scattering, it has to be
expected that scalar mesonic resonances must be the domain to search for
them \cite{ZPC30p615}.

For $K\pi$ and $\pi\pi$ scattering, threshold is far below the ground states
of the respective confinement spectra $u\bar{s}$ and
$\left( u\bar{u}+d\bar{d}\right) /\sqrt{2}$,
which are at 1.389 GeV resp.\ 1.287 GeV in our model
with the parameters of \cite{PRD27p1527}.
This is the situation comparable to Fig.~\ref{extrapole}a.
Consequently, the poles associated with the $K^{\ast}_{0}(791)$ and
the $f_{0}$(400--600) resonances are still deep in the second Riemann sheet,
causing broad structures in the scattering cross sections.
In Ref.~\cite{ZPC30p615}, the respective poles were found at $(727\!-\!i\,263)$
MeV and $(470\!-\!i\,208)$ MeV, well explaining the scattering data.

The $K\bar{K}$ threshold at some 0.99 GeV is much closer to the
$s\bar{s}$ and $\left( u\bar{u}-d\bar{d}\right) /\sqrt{2}$
ground states at 1.491 GeV and 1.287 GeV.
The poles end up just below threshold, as bound states of the coupled systems
\cite{HEPPH0304105,HEPPH0207022}.
But $s\bar{s}$ also couples to $\pi\pi$ in our many-channel model
\cite{ZPC30p615}, through the chain

\begin{equation}
s\bar{s}\;\;\longrightarrow\;\;
K\bar{K}\;\;\longrightarrow\;\;
\fnd{1}{\sqrt{2}}\;\left( u\bar{u}+d\bar{d}\right)\;\;\longrightarrow\;\;
\pi\pi
\;\;\; ,
\label{ssbarpipi}
\end{equation}

\noindent
whereas $\left( u\bar{u}-d\bar{d}\right) /\sqrt{2}$ also couples to $\pi\eta$.

The $s\bar{s}$ channel couples to $\pi\pi$ through the chain
(\ref{ssbarpipi}), implying that the effective coupling is small
\cite{PLB559p49,PLB521p15}.
Hence, the resulting pole displacement relative to the mentioned
bound state below threshold, is small.
In \cite{ZPC30p615} one finds $(994\!-\!i\,20)$ MeV for this pole,
associated to the $f_{0}(980)$ resonance.

In the case of the $a_{0}(980)$ resonance, both $K\bar{K}$ and $\pi\eta$
couple directly, through OZI-allowed pair creation/annihilation, to
$\left( u\bar{u}-d\bar{d}\right) /\sqrt{2}$. However, the coupling intensity
for $\pi\eta$ is three times smaller than for $K\bar{K}$.
Consequently, we may deal with $\pi\eta$ as a perturbative correction
to $K\bar{K}$ here, resulting in an additional shift of the pole
in the negative-imaginary direction in the complex-energy plane.
In the full multi-channel calculation of Ref.~\cite{ZPC30p615}, the pole
corresponding to the $a_{0}(980)$ was obtained at $(968\!-\!i\,28)$~MeV.

\section{The model's K matrix}

In its one but simplest form, the model's $K$ matrix for low-energy elastic
meson-meson scattering in $\ell$ wave is given by the expression

\begin{equation}
K_{\ell}(p)\; =\;
\fnd{2a^{4}\lambda^{2}\mu p\;
j^{2}_{\ell}(pa)\;
\dissum{n=0}{\infty}
\fnd{\abs{{\cal F}_{n\ell_{c}}(a)}^{2}}
{E(p)-E_{n\ell_{c}}}}
{2a^{4}\lambda^{2}\mu p\;
j_{\ell}(pa)\; n_{\ell}(pa)\;
\dissum{n=0}{\infty}
\fnd{\abs{{\cal F}_{n\ell_{c}}(a)}^{2}}
{E(p)-E_{n\ell_{c}}}\; -\; 1}
\;\;\; .
\label{partialKmatrix}
\end{equation}

In formula~(\ref{partialKmatrix}), $p$, $\mu$, and $E(p)$ respectively
represent the linear momentum, reduced mass, and total invariant mass of the
two-meson system. Furthermore, $a$ stands for the average distance where
quark-pair creation/annihilation takes place, ${\cal F}_{n\ell_{c}}$ is the
radial part of the eigensolution of the confinement system with radial
excitation $n$ in $\ell_{c}$ wave and for eigenvalue $E_{n\ell_{c}}$,
$\,j_{\ell}$ and $n_{\ell}$ are the spherical Bessel resp.\ Neumann funtions,
and $\lambda$ represents the coupling between the two-meson and the confinement
systems. In Ref.~\cite{syracuse} we give a derivation of formula
(\ref{partialKmatrix}).

One easily deduces for small $\lambda$,
assuming there exists a pole close to one of the confinement energy
eigenvalues, say $E_{\nu\ell_{c}}$, that the pole displacement,
defined by

\begin{displaymath}
\Delta_{\nu\ell_{c}}\; =\; E\; -\; E_{\nu\ell_{c}}
\;\;\; ,
\end{displaymath}

\noindent
is given by the relation

\begin{equation}
\Delta_{\nu\ell_{c}}\;\approx\;
2a^{4}\lambda^{2}\mu p^{(\nu)}\;
j_{\ell}\left( p^{(\nu)}a\right)\;\left[
n_{\ell}\left( p^{(\nu)}a\right)\; -\;
i\, j_{\ell}\left( p^{(\nu)}a\right)\;\right]\;
\abs{{\cal F}_{\nu\ell_{c}}(a)}^{2}
\;\;\; ,
\label{shift}
\end{equation}

\noindent
where $p^{(\nu)}$ is the linear momentum corresponding to the
energy $E_{\nu\ell_{c}}$.

Notice from relation~(\ref{shift}) that the imaginary part of the
pole displacement is negative, as it should be when $E_{\nu\ell_{c}}$
is above threshold.

It should also be clear at this stage that, by letting $\lambda$ increase,
one may follow the pole's trajectory, but when next $\lambda$ is decreased,
the pole nicely returns to $E_{\nu\ell_{c}}$.
Hence, the extra poles do not show up this way.

In the other limit of very large $\lambda$, one ends up with
a very simple relation for the pole positions, namely

\begin{equation}
j_{\ell}\left( pa\right)\; +\; i\,
n_{\ell}\left( pa\right)\; =\; 0
\;\;\; ,
\label{hardsphere}
\end{equation}

\noindent
which is the relation for infinite-hard-sphere scattering in $\ell$ wave.

For data analysis of meson-meson scattering data, the form of the $K$ matrix in
Eq.~(\ref{partialKmatrix}) is as good as any Breit-Wigner expansion.
But it has two advantages:
\vspace{0.2cm}

\begin{itemize}
\item
It automatically incorporates the extra poles, without any additional
parameter.
\item
In the limit $\lambda\downarrow 0$, one obtains the confinement spectrum,
equivalent to quenched-lattice spectra.
\end{itemize}
\vspace{0.2cm}

In real data analysis, one may just take a few terms of the summation over
the radial quantum number $n$, and approximate the rest of the summation
by a constant.
Since, furthermore, confinement is not understood,
the moduli squared of the eigenstates $\abs{{\cal F}_{n\ell_{c}}(a)}^{2}$
and the eigenvalues $E_{n\ell_{c}}$
turn into fit parameters for data analysis, to be adjusted to experiment.
The resulting {\it experimentally} \/determined confinement spectrum
must coincide with the quenched-lattice spectra.
This moreover implies that the light scalars are no issue for
quenched-lattice calculations.
Thus, a perfect mediator between experiment and low-energy QCD 
is created through expression (\ref{partialKmatrix}).

Finally, note that the expression~(\ref{partialKmatrix}) could as well
serve for baryon-meson \cite{HEPPH0305292} or baryon-antibaryon scattering.

\section{Conclusions}

Through unitarization, transition matrices can be constructed
describing meson-meson scattering.
Here we discussed a method which reproduces low-energy data
for those two-meson systems that couple through OZI-allowed processes
to nonexotic confined states. By studying its pole structure,
we discovered that there exist two types of singularities
in the scattering amplitude when analytically continued to
complex energies.
One type of poles can be directly related to the confinement spectrum,
even for cases where pole displacements are large.
In the limit of large coupling, these poles are equivalent to scattering
from a hard sphere. Poles of the other type stem from the background, and are
mainly important for scattering in $S$ wave, the light scalar mesons being
their most important manifestation, as well as the recently observed \babar\
\cite{HEPEX0304021} resonance \cite{HEPPH0305035,syracuse}.

When applied in data analysis, the model's $K$ matrix might build
the bridge between experiment and QCD lattice calculations.
\vspace{0.3cm}

{\bf Acknowledgments}:
One of us (EvB) wishes to thank the organizers of the Scalar Mesons Workshop
for kindly inviting him and for their warm hospitality.
It was a pleasure to be hosted at the beautiful surroundings and facilities
of SUNY Institute of Technology at Utica, New York.
He, moreover, thanks Carla G\"{o}bel, Sandra Malvezzi, Luigi Moroni,
Brian Meadows, Ign\'{a}cio Bediaga, and Robert Jaffe for useful suggestions
for future work.
This work was partly supported by the
{\it Funda\c{c}\~{a}o para a Ci\^{e}ncia e a Tecnologia}
\/of the {\it Minist\'{e}rio da
Ci\^{e}ncia e da Tecnologia} \/of Portugal,
under contract number
POCTI/\-FNU/\-49555/\-2002.

\end{document}